# Recently synthesized $(Ti_{1-x}Mo_x)_2AlC$ ($0 \leq x \leq 0.20$) solids solutions: Deciphering the structural, electronic, mechanical and thermodynamic properties via ab initio simulations

M. A. Ali[a,*], S. H. Naqib[b,**]

[a]Department of Physics, Chittagong University of Engineering and Technology (CUET), Chattogram 4349, Bangladesh
[b]Department of Physics, University of Rajshahi, Rajshahi-6205, Bangladesh

ABSTRACT

The structural, electronic, mechanical and thermodynamic properties of $(Ti_{1-x}Mo_x)_2AlC$ ($0 \leq x \leq 0.20$) were explored via density functional theory. The obtained lattice constants agree well with the experimental values. The electronic band structure confirms the metallic nature. Strengthening of covalent bonds due to Mo substitution is confirmed from the study of band structure, electronic density of states and charge density mapping. The elastic constants satisfy the mechanical stability criteria. Strengthening of covalent bonds leads to enhanced mechanical properties. $(Ti_{1-x}Mo_x)_2AlC$ compounds are found to exhibit brittle behavior. The anisotropic nature of $(Ti_{1-x}Mo_x)_2AlC$ is revealed from the direction dependent Young's modulus, compressibility, shear modulus and Poisson's ratio as well as the shear anisotropic constants and the universal anisotropic factor. The Debye temperature, minimum thermal conductivity, Grüneisen parameter and melting temperature of $(Ti_{1-x}Mo_x)_2AlC$ have been calculated for different Mo contents. Our calculated values are compared with reported values, where available.

Keywords: $(Ti_{1-x}Mo_x)_2AlC$ solid solutions; Electronic band structure; Mechanical properties; Thermodynamic properties; Density functional theory

Corresponding authors: *ashrafphy31@cuet.ac.bd; **salehnaqib@yahoo.com

## 1. Introduction

Intrinsically nanolaminated MAX phase compounds have drawn significant attention of the materials science community because of their unusual combination of physical properties such as damage tolerance, high stiffness, resistance to corrosion and thermal shock like ceramics as well as fairly high electrical and thermal conductivity, and machinability like metallic compounds [1, 2]. This outstanding combination of these properties makes MAX compounds suitable for extensive use in a number of applications mentioned in literatures [1–6]. In 1960s, the powder forms of these carbides/nitrides were first discovered [7–10], but the interest has been reinvigorated after the studies of Barsoum and coworkers in the 1990s [6, 11] which disclosed their interesting properties in details for the first time. Physical properties of these systems originate mainly due to the layered structure of MAX phases that have A-layers (Sn, Al, Ge, etc.) sandwiched in between $M_{n+1}X_n$ sheets (e.g., $Ti_2C$) with the stacking of atomic layers along the $c$-axis [12, 13]. Till date, a number of compounds (around 80) belonging to MAX phases have already been synthesized [1, 14], theoretically around 155 are known [15] out of a possible 665 viable MAX phases ($M_{n+1}AX_n$, n = 1–4) [16] and the number of reports performed theoretically as well as experimentally are increasing tremendously [3, 17-38]. Study of existing and potentially viable MAX phase nanolaminates has established itself as an important subfield of materials science research.

The $Ti_2AlC$ compound belongs to the Ti-Al-C system representing a large family of solids with overall advantages of MAX phases and thus becomes one of widely studied compounds from the 211 MAX phase materials [39]. The M-Al-X phases are most attractive for high-temperature applications, such as $Ti_2AlC$ and $Cr_2AlC$ because of their high oxidation resistant (up to 1400 ºC) resulted from the slow growing protective alumina ($Al_2O_3$) layers [40, 41]. The coefficient of thermal expansion (8.2 $\times 10^{-6}$ $K^{-1}$) of $Ti_2AlC$ [42] is almost equal to that of α-$Al_2O_3$ (9.3$\times 10^{-6}$ $K^{-1}$) [43], gives it another advantage by making it resistant to thermal stresses upon thermal cycling [44]. However, low hardness and strength limit its potential use as a structural component like other binary carbide ceramics [45]. Therefore, an increase of hardness and bonding strength along with melting temperature of $Ti_2AlC$ will be highly advantageous to extend its domain of applications.

Scientific community are always keen on the reporting of new MAX phase solid solutions, one of the efficient ways of altering the properties of existing MAX phase materials, opening new wing of research and widening their field of applications. Enhancement of many important properties such as oxidation [46-50], fracture toughness [51, 52], strength [48, 51] and self-healing properties [47, 53] are reported in the solid solutions. Attempts have also been made to increase the hardness and strength of $Ti_2AlC$ both theoretically and experimentally [42, 45, 54-57]. A considerable increase in elastic modulus is reported by Nb substitution in $Ti_2AlC$ [42]. Strengthening of $M_xM'_{1-x}AlC$ (M and M′ = Ti, V, and Cr) solid solutions has been reported by Wang et al. [56]. Meng et al. [57] have investigated $Ti_2AlC$ and $(Ti_{0.8}V_{0.2})_2AlC$ where enhancement of Vickers hardness, flexural strength, and shear strength were reported due to V substitution. A considerable improvement in compressive strength has already been reported for $Ti_2AlC_{1-x}N_x$ [58]. Recently, Pan et al. [59] have successfully synthesized the Mo substituted $Ti_2AlC$ solid solution where a significant increase of Vickers hardness, flexural strength, and fracture toughness were reported. The atom Mo has the possibility to form $(Ti_{1-x}Mo_x)_2AlC$ solids solutions because of its smaller atomic radius compared to Ti. The earlier reports also proved the ability of Mo to modify the microstructural and mechanical properties of MAX phase materials [45, 60]. Moreover, we have investigated the effect of M mixing on the physical properties of $(Zr,Ti)_2AlC$ in our previous study [27]. Thus, the mechanical and thermodynamic properties are expected to be upgraded by substituting Mo into $Ti_2AlC$.

Therefore, we aimed to study the structural, electronic, mechanical and thermodynamic properties of $(Ti_{1-x}Mo_x)_2AlC$ ($0 \leq x \leq 0.20$) solids solutions theoretically for the first time where the modification of mechanical and thermodynamic properties due to alloying with Mo are explored in details. The value of $x$ is limited up to 0.20 and only the experimentally synthesized compositions are chosen to keep away from the controversy of the thermodynamical stability.

The rest of the paper has been structured as follows. Section 2 describes the computational methodology in brief. Results are presented, analyzed and discussed in Section 3. Finally, important conclusions from this study are summarized in Section 4.

## 2. Calculation methods

The CAmbridge Sequential Total Energy Package (CASTEP) code [61] was used to calculate the physical properties of $(Ti_{1-x}Mo_x)_2AlC$ solids solutions via the density functional theory (DFT) [62, 63] based on the plane-wave pseudopotential method. The generalized gradient approximation (GGA) of Perdew–Burke–Ernzerhof (PBE) is used as an exchange and correlation functional [64]. The pseudo atomic calculation was performed for C - $2s^2\ 2p^2$, Al - $3s^2\,3p^1$, Ti - $3s^2\,3p^6\,3d^2\,4s^2$ and Mo - $4s^2\,4p^6\,4d^5\,5s^1$ electronic orbitals. The cutoff energy of 400 eV and a k-point mesh of 12×12×3 [65] were set for good convergence using the Broyden Fletcher Goldfarb Shanno (BFGS) geometry optimization [66] technique. The self-consistent convergence of the total energy was $5 \times 10^{-6}$ eV/atom, and the maximum force on the atom was 0.01 eV/Å. The maximum ionic displacement was set to $5 \times 10^{-4}$ Å, and a maximum stress of 0.02 GPa was used. The Virtual Crystal Approximation (VCA) approach was employed to model the solid solutions. A number of works have been published recently using the VCA [27, 28, 67-70] which resulted in reliable estimates of thermo-physical and electronic band structure properties of MAX phase compounds.

### 3. Results and discussion

### 3.1. *Structural properties*

The MAX phases are crystallized in hexagonal structure with the space group $P6_3/mmc$. There are 8 atoms (for 211 system) in the unit cell of which 4 *M*, 2 *A* and 2 *X* atoms are contributed to the total 8 atoms. In case of $Ti_2AlC$, the position of carbon atoms is 2a, Ti is 4f and Al atoms is 2c Wyckoff positions, as shown schematically in Fig. 1(a). In order to calculate the ground state physical properties, we have to obtain the equilibrium structure by optimizing the total energy of the solids under consideration. Optimized lattice constants are presented in Table 1 together with values obtained by others [71-75].

**Table 1**
The obtained lattice parameters (*a* and *c*), *c/a* ratio, and cell volumes (*V*) for the $(Ti_{1-x}Mo_x)_2AlC$ ($0 \le x \le 0.20$) solid solutions, together with values obtained by previous studies.

| *x* | *a* (Å) | *c* (Å) | *c/a* | *V* (Å³) |
|---|---|---|---|---|
| 0.00 | 3.067 | 13.737 | 4.478 | 111.94 |
| | 3.058 [a] | 13.642 [a] | 4.461 [a] | 106.62 [a] |
| | 3.066 [b] | 13.694 [b] | 4.466 [b] | 111.49 [b] |
| | 3.04 [c] | 13.60 [c] | | |
| | 3.053 [d] | 13.640 [d] | 4.468 [d] | |
| | 3.06 [e] | 13.67 [e] | 4.46 [e] | |
| 0.05 | 3.0620 | 13.677 | 4.466 | 111.06 |
| 0.10 | 3.0552 | 13.636 | 4.463 | 110.23 |
| 0.15 | 3.0461 | 13.618 | 4.470 | 109.43 |
| 0.20 | 3.0334 | 13.606 | 4.485 | 108.66 |

[a]Expt.[71], [b]GGA [72], [c]LDA [73], [d]GGA [74], [e]GGA [75]

Our calculated lattice constant of Ti$_2$AlC: $a$ and $c$ are just 0.29% and 0.69%, respectively, higher than the experimentally obtained values. This ensures very high level of consistency of present first-principles calculations. Fig. 1(b) illustrates the variation of $a$ and $c$ with Mo contents from which it is clear that the lattice constants $a$ and $c$ decrease with Mo content. The decrement of lattice constants can be largely attributed to the difference in ionic radii of Mo (0.69 Å) and Ti (0.74 Å). Since Mo is smaller than the Ti, its substitution causes shrinkage of the unit cell. The decrease of lattice constant $c$ with Mo content is reported in experiment while the values of $a$ have been found to vary by little [59]. However, the compositions considered here roughly follow the Vegard's law [76].

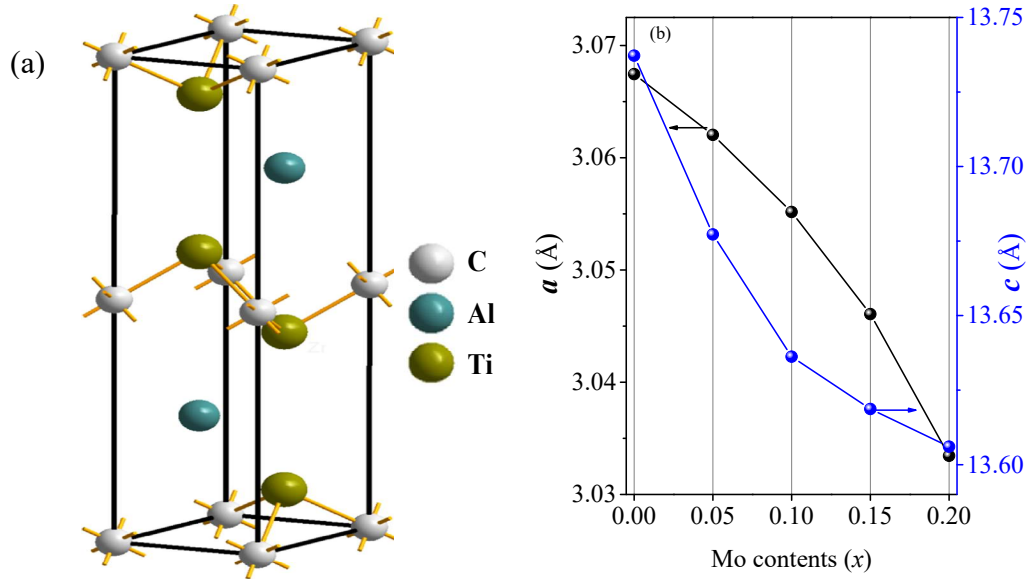

Fig. 1. (a) Crystal structure (unit cell) of the Ti$_2$AlC compound and (b) lattice constants $a$ (Å) and $c$ (Å) for different Mo contents.

### 3.2. Electronic properties

Fig. 2(a) shows the electronic band structure of Ti$_2$AlC in which no band gap is observed between valence and conduction bands and the density of states (DOS) at Fermi level is 3.04 states/eV [Fig. 2(c)]. Accordingly, Ti$_2$AlC is found to be metallic in nature which is able to conduct electrically as well as thermally, which is in good accord with the experimental results of Barsoum et al. [77]. The electrical conductivity exhibited by Ti$_2$AlC is direction dependent. The paths Γ-A, H-K and M-L are along the $c$-direction which exhibited much less degree of energy dispersion. On the other hand, the paths A-H, K-Γ, Γ-M and L-H are in the basal planes, exhibiting much greater degree of energy dispersion [73]. This shows that the electronic effective mass tensor is higher for $c$-axis conduction compared to that in the $ab$-plane. Therefore, anisotropic electrical conductivity is predicted for Ti$_2$AlC i.e., the electrical conductivity is higher in the basal plane than that along the *c*-direction. Fig. 2(b) shows the band structure of $(Ti_{0.95}Mo_{0.05})_2AlC$ exhibiting almost similar characteristics except that the position of some of the dispersion curves. These curves are observed to be shifted slightly to lower energy. The band structures (not shown) for other compositions (*x* = 0.10, 0.15 and 0.20) also show similar nature with shifting of the curves towards lower energy which increases with increasing Mo contents as can be observed from Fig. 2(d).

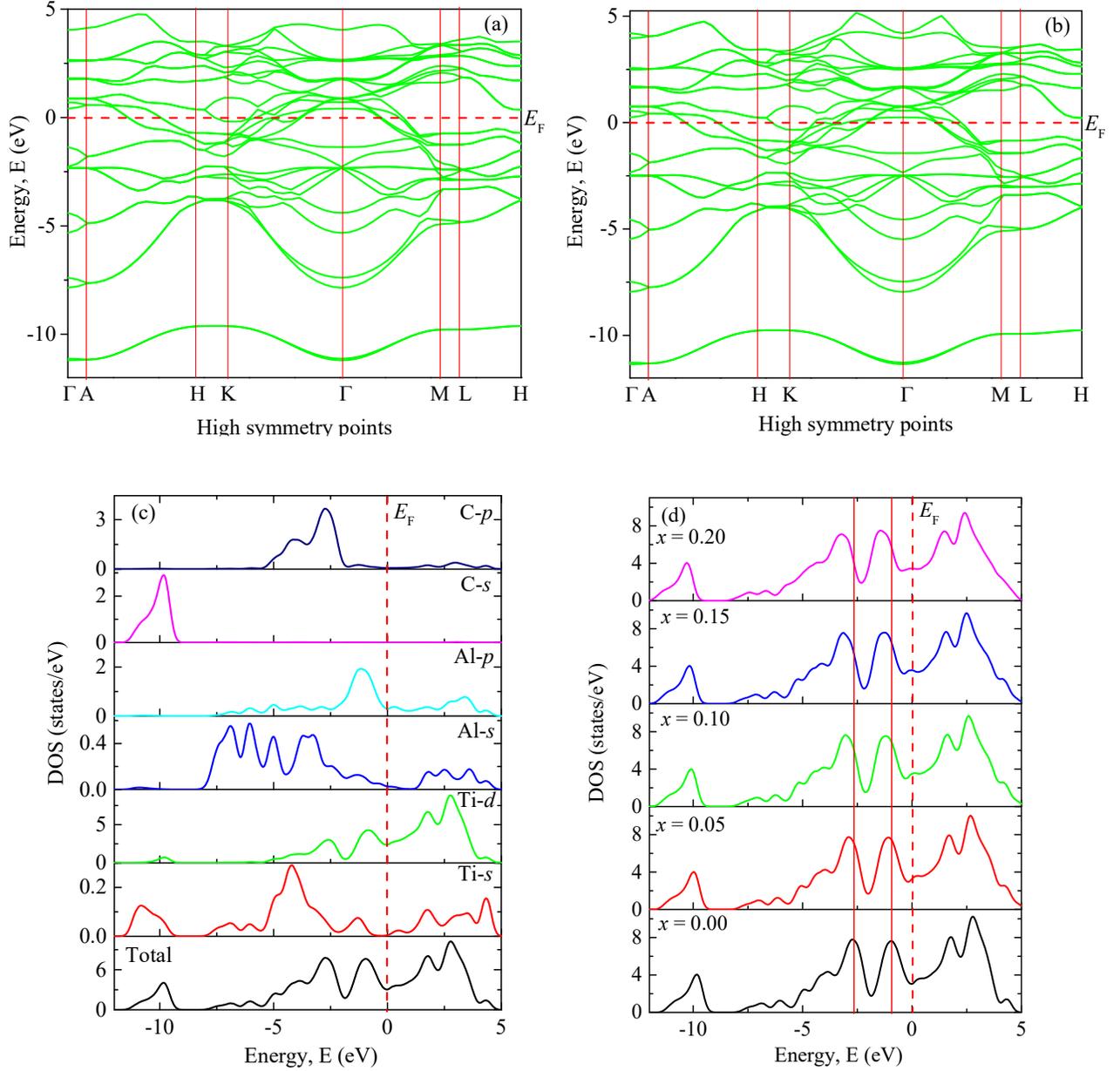

Fig. 2. The electronic band structure of (a) $Ti_2AlC$ and (b) $(Ti_{0.95}Mo_{0.05})_2AlC$. Total and partial DOS of (c) $Ti_2AlC$ and (d) the total DOS for different Mo contents.

Fig. 2(c) shows total and partial DOS of $Ti_2AlC$ which gives better understanding of bonding characteristics in the system. The partial DOS is used to disclose the contribution from different electronic states. The dominating contribution comes from Ti-3$d$ with a very small contribution from Al 3$p$ orbital at the Fermi level. The hybridization among different states is observed in different energy ranges. The peak in the energy range -9 eV to -11.5 eV comes from hybridization of C-2$s$ and Ti-4$s$ states with a prime contribution from C-2$s$ states. The peak in the range -1.8 eV to -3.6 eV comes from the hybridization of C-2$p$ and Ti-3$d$ states. This bonding is primarily covalent in nature resulting from strong mixing between C-2$p$ and Ti-3$d$ states which dominates the higher energy range. The peak just below the Fermi level is attributed from the hybridization of Ti-3$d$ and Al-3$p$ states that results in a comparatively weaker covalent bonding than the previous one [78]. The energy range 0 to 5 eV is dominating by the metal-to-metal $dd$ interactions and anti-bonding states. The covalent bond among the unlike atoms determines the overall bonding strength and consequently the mechanical properties of the MAX phases. The total and partial DOS of Mo substituted compositions are also calculated (not shown) and revealed the same qualitative nature of bonding except the positions of the peaks (energy range of hybridization) that comes from the contribution of Mo-4$d$ sates. The shifting of these peaks can be visualized in Fig. 2(d). Fig. 2(d) shows the total DOS for different Mo contents in which two vertical red solid lines are drawn to refer the position of the peaks for the $x$ = 0 compound. These two reference lines clearly demonstrate the shifting of the peaks towards lower energy with increasing Mo content. The hybridized energy states always give an indication of the bonding strength. In general, lower the energy states stronger the bonding strength. Therefore, lowering of the position of the peaks indicates the better hybridization among different states and stronger covalent bonds between them. The strengthening of the covalent bonds can also be observed from Fig. 3. Fig. 3 shows the charge density mapping (CDM) (in the units of e/$\text{Å}^3$) in the (101) crystallographic plane. CDM is an important tool to describe the bonding nature of solids. The occupation positions of atoms have been labeled in the figure. A very clear change is observed in the charge density at the Ti/Mo positions with varying level of Mo substitution. It is quite clear from Fig. 3 that the accumulation of charges at the Ti sites increased to a considerable extent owing to Mo substitution and it increases gradually with increasing Mo contents. Therefore, the strength of covalent bond is expected to increase because of the better mixing of $d$ and $p$ electronic states. Hence, the dominating covalent bond between Ti/Mo-*d* and C-*p* states as well as comparatively weaker covalent bond between Ti/Mo-*d* and Al-*p* states is assumed to be strengthened. Hence, it is reasonable to expect that the overall bonding strength within (Ti$_{1-x}$Mo$_x$)$_2$AlC ($0 \leq x \leq 0.20$) solid solutions will increase owing to Mo substitution and this should be reflected in the enhanced mechanical properties presented in the following section.

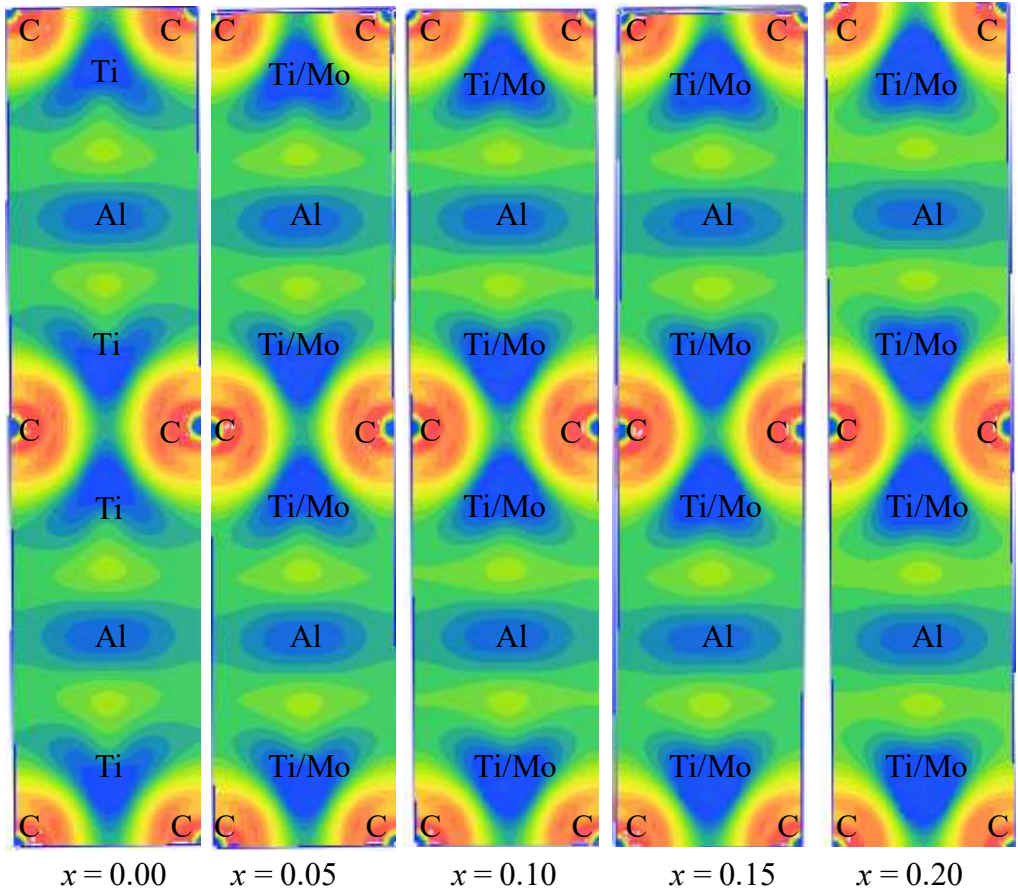

Fig. 3. The electronic charge density mapping for different Mo contents.

### 3.3. *Mechanical properties*

The knowledge of elastic properties e.g. stiffness constant, elastic moduli, elastic anisotropy factors and ductile or brittle characteristic is of scientific interest for practical use of engineering materials. The elastic constants studied here are obtained from linear finite strain-stress method as implemented in the CASTEP [79]. The stiffness constants ($C_{ij}$) of (Ti$_{1-x}$Mo$_x$)$_2$AlC ($0 \leq x \leq 0.20$) solid solutions are calculated and presented in Table 2. The confirmation of mechanical stability is essential for studying the mechanical properties of a solid and we have checked the mechanical stability conditions of $(Ti_{1-x}Mo_x)_2AlC$  ($0 \leq x \leq 0.20$) solid solutions using the Born criteria [80] for hexagonal system: $C_{11} > 0$, $C_{11}-C_{12} > 0$, $C_{44} > 0$, $(C_{11} +C_{12})C_{33} - 2C_{13} > 0$. To judge the accuracy of our calculations, the reported elastic constant values of $Ti_2AlC$ are also presented herein. Comparison with the calculated values presented here ensures the reliability of our theoretical estimations.

**Table 2**

The elastic constants $C_{ij}$ (GPa), bulk modulus, $B$ (GPa), shear modulus, $G$ (GPa), Young's modulus, $Y$ (GPa), Poisson ratio, $v$, Pugh ratio, $G/B$ and Cauchy pressure of $(Ti_{1-x}Mo_x)_2AlC$  ($0 \leq x \leq 0.20$) solid solutions.

| $x$ | $C_{11}$ | $C_{12}$ | $C_{13}$ | $C_{33}$ | $C_{44}$ | $B$ | $G$ | $Y$ | $v$ | $G/B$ | Cauchy Pressure | *Ref.* |
|---|---|---|---|---|---|---|---|---|---|---|---|---|
|      | 302 | 59 | 58 | 269 | 107 | 135 | 113 | 265 | 0.17 | 0.84 | -48 | This study |
|      | 308 | 55 | 60 | 270 | 111 | 137 | 118 |     |      | 0.86* |     | Ref. [56, 74] |
| 0.00 | 309 | 72 | 68 | 284 | 112 | 146 | 115 | 273 | 0.19 | 0.78* |     | Ref. [72] |
|      | 301 | 69 | 62 | 266 | 106 | 139 |     |     |      |       |     | Ref. [81] |
|      | 306 | 59 | 59 | 272 | 109 | 138 |     |     |      |       |     | Ref. [82] |
|      | 302 | 68 | 64 | 268 | 107 | 141 |     | 111 | 264 | 0.18 | 0.79* | Ref. [83] |
| 0.05 | 304 | 68 | 66 | 277 | 116 | 143 | 116 | 274 | 0.18 | 0.81 | -49 |    |
| 0.10 | 307 | 55 | 73 | 280 | 122 | 145 | 119 | 280 | 0.18 | 0.82 | -67 |    |
| 0.15 | 304 | 54 | 84 | 272 | 128 | 148 | 120 | 283 | 0.18 | 0.81 | -74 |    |
| 0.20 | 323 | 62 | 83 | 278 | 131 | 154 | 124 | 293 | 0.18 | 0.81 | -69 |    |

*Calculated using published data.

Additionally, the uniaxial directional elastic behaviors can also be understood from the elastic constants. Like the usual trends in MAX phases, the values of $C_{11}$ are observed to be higher than $C_{33}$ for $(Ti_{1-x}Mo_x)_2AlC$ solid solutions. The $C_{11}$ and $C_{33}$ measure the resistance required to compress the compound along [100] (along $a$ – axis) direction and along [001] (along $c$ - axis) direction, respectively. The higher values of $C_{11}$ and $C_{33}$ are found in Mo substituted compositions revealing the increased bonding strength both along $a$ and $c$ directions. Another important stiffness constant $C_{44}$ also known as shear stiffness, by itself corresponds to a change in shape keeping the volume unchanged and provides with the information regarding hardness directly [84]. The $C_{44}$ is also found to be increased by 8.4%, 14.0%, 19.6% and 22.4% for $x =$ 0.05, 0.10, 0.15 and 0.20, respectively. The obtained values of $C_{13}$ and $C_{12}$ are close to each other as usually observed in MAX phases [21].

The mechanical parameters such as bulk moduli ($B$), shear moduli ($G$), Young's moduli (Y), Poisson's ratio ($v$) are calculated using the formalism that can be found elsewhere [85]. The calculated elastic moduli are found in good agreement with the reported values as shown in

Table 2. The two main mechanical parameters of polycrystalline solids are bulk and shear moduli. The ability to resist volume deformation and plastic deformation is represented by $B$ and $G$ which are found to be increased by 5.9%, 7.4%, 10.0%, 14.1% and 2.6%, 5.3%, 6.2%, 9.7% for $x = 0.05$, 0.10, 0.15 and 0.20, respectively. Another important mechanical parameter for polycrystalline solids is the Young's modulus, by itself representing the stiffness of solids where the higher value indicates the stiffer nature and vice versa. The value of $Y$ is found to be increased by 3.3 %, 5.6%, 6.8%, 10.6% for $x = 0.05$, 0.10, 0.15 and 0.20.

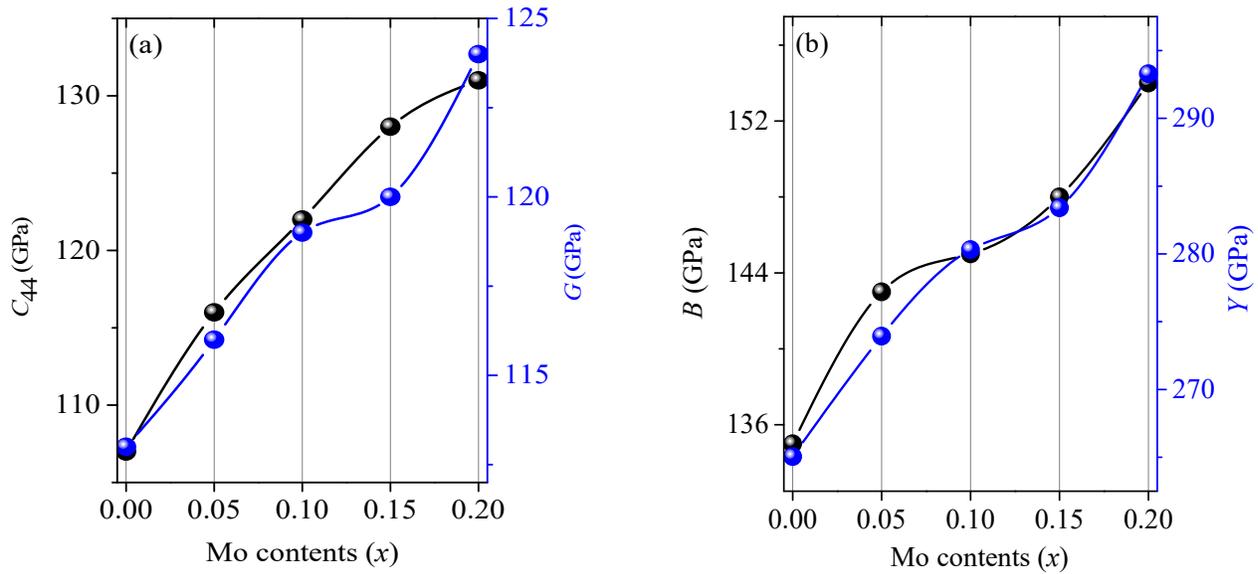

Fig. 4. The variation of $C_{44}$, $B$, $G$ and $Y$ with Mo contents in $(Ti_{1-x}Mo_x)_2AlC$ solid solutions.

As we have already discussed that the direct information regarding hardness can be known from $C_{44}$ which is found to be increased by 8.4%, 14.0%, 19.6% and 22.4% for $x = 0.05$, 0.10, 0.15 and 0.20, respectively. The enhancements of mechanical parameters $B$, $G$ and $Y$ are considerable due to alloying with Mo and are illustrated in Fig. 4. Although, these parameters cannot predict directly the actual hardness of solids, but all these elastic moduli are usually higher for comparatively harder solids [86]. It should be noted that Pan *et al*. [59] have measured the Vickers hardness, flexural strength and fracture toughness and found those to be increased with Mo contents with a maximum increase of 44, 34 and 136%, respectively for $(Ti_{0.80}Mo_{0.20})_2AlC$.

*The brittleness of $(Ti_{1-x}Mo_x)_2AlC$  ($0 \leq x \leq 0.20$)*

The brittleness or ductility of the studied materials has been assessed by calculating the Pugh ratio ($G/B$), Poisson's ratio ($v$) and Cauchy pressure as presented in Table 2. The ratio $G/B$ is named after the name of Pugh, a widely used parameter of brittleness or ductility of solids. A $G/B$ value of 0.571 indicates the border line between the brittleness and ductility [87]. Another widely used parameter to predict brittleness or ductility is the Poisson's ratio where the value 0.26 indicates the border line between the brittleness or ductility as proposed by Frantsevich *et al*. [88]. A value of $G/B$ greater than 0.571 (or $v$ lower than 0.26) indicates the brittle character of solids while a value of $G/B$ lower than 0.571 (of $v$ greater than 0.26) exhibits the ductile character of the solids. The values presented in Table 2 revealed the brittle character of the studied solid solutions. The difference between two stiffness constants ($C_{12}$ - $C_{44}$) is known as the Cauchy pressure ($CP$). This parameter is assumed to define the bonding nature in solids depending on the value of $CP$ (either positive or negative) [89]. For ductile metallic material the values of $CP$ is positive while a negative value points toward the covalent bonding with brittle nature with directional bonding character. Furthermore, the magnitudes of negative values exhibit the level of dominance of covalent bonding in the overall makeup of the solid. The large negative values $CP$ for $(Ti_{1-x}Mo_x)_2AlC$ solid solutions revealed the dominant role of directional covalent bonding in association with brittle nature within the studied compounds. The $CP$ increases with elevated level of Mo implying that Mo substitution makes covalent contribution stronger in $(Ti_{1-x}Mo_x)_2AlC$.

*The elastic anisotropy*

The compositions considered here are mechanically anisotropic. The study of anisotropy has been considered because of its involvement with the many physical processes of engineering importance, i.e., the creation of plastic deformation in solids, enhanced charged defect mobility, dynamics of micro-scale cracking in ceramics etc. [90].

The values of $Y$, $B$, $G$ and $v$ are direction depend that can be presented by three-dimensional (3D) plots. The plotting of 3D contour graphs is an attractive way of presenting the mechanical anisotropy where a perfect sphere designates perfect isotropy of solids and departure from it spherical shape measures the level of anisotropy [91]. The 3D plots of $Y$, $B$ (in terms of compressibility), $G$ and $v$ of $(Ti_{1-x}Mo_x)_2AlC$ solid solutions obtained using the ELATE program [92] are shown in Figs. (5-8). These plots were generated from the calculated values of $C_{ij}$ of $(Ti_{1-x}Mo_x)_2AlC$. The anisotropy level of the studied materials can be understood from the figures where the most anisotropic composition is $(Ti_{0.85}Mo_{0.15})_2AlC$. Various anisotropy indices are also presented in Table 3 where the value of $A$ is 1 for isotropic solids which represents a perfect sphere in the ELATE generated profile.

**Table 3**

The minimum and the maximum values of the Young's modulus, 1/bulk modulus, shear modulus, Poisson's ratio and anisotropy indices of $(Ti_{1-x}Mo_x)_2AlC$ solid solutions.

| $x$ | $Y_{min.}$ (GPa) | $Y_{max.}$ (GPa) | $A$ | $(1/B)_{min}$ $(TPa^{-1})$ | $(1/B)_{max}$ $(TPa^{-1})$ | $A$ | $G_{min.}$ (GPa) | $G_{max.}$ (GPa) | $A$ | $v_{min.}$ (GPa) | $v_{max.}$ (GPa) | $A$ |
|---|---|---|---|---|---|---|---|---|---|---|---|---|
| 0.00 | 251 | 282 | 1.12 | 2.3 | 2.7 | 1.15 | 108 | 121 | 1.12 | 0.15 | 0.19 | 1.25 |
| 0.05 | 254 | 279 | 1.10 | 2.2 | 2.5 | 1.13 | 112 | 118 | 1.05 | 0.16 | 0.19 | 1.16 |
| 0.10 | 250 | 289 | 1.15 | 2.3 | 2.4 | 1.04 | 109 | 125 | 1.14 | 0.13 | 0.22 | 1.78 |
| 0.15 | 233 | 296 | 1.27 | 2.2 | 2.2 | 1.01 | 100 | 128 | 1.27 | 0.10 | 0.27 | 2.69 |
| 0.20 | 242 | 306 | 1.26 | 2.0 | 2.3 | 1.12 | 107 | 131 | 1.22 | 0.12 | 0.26 | 2.06 |

The three shear anisotropic factors $A_1$, $A_2$ and $A_3$ correspond to three independent elastic shear constants for hexagonal crystals and are calculated using the formalism that can be found elsewhere [28]. The values of $A_i$'s are not 1 (for isotropic solids, $A_i = 1$), revealed the anisotropic nature of the $(Ti_{1-x}Mo_x)_2AlC$ solid solutions. The universal anisotropic index $A^U$ [$A^U = 5\frac{G_V}{G_R} + \frac{B_V}{B_R} - 6 \geq 0$] of $(Ti_{1-x}Mo_x)_2AlC$ solid solutions are also calculated and found to have non-zero value, revealing the anisotropic nature of the $(Ti_{1-x}Mo_x)_2AlC$ solid solutions [93].

**Table 4**

Anisotropic factors, $A_1$, $A_2$, $A_3$, and universal anisotropic index $A^U$ of $(Ti_{1-x}Mo_x)_2AlC$ solid solutions.

| $x$ | $A_1$ | $A_2$ | $A_3$ | $A^U$ |
|---|---|---|---|---|
| 0.00 | 1.04 | 0.88 | 0.91 | 0.22 |
| 0.05 | 0.95 | 0.98 | 0.94 | 0.23 |
| 0.10 | 0.86 | 0.97 | 0.83 | 0.19 |
| 0.15 | 0.74 | 1.02 | 0.75 | 0.19 |
| 0.20 | 0.77 | 1.00 | 0.78 | 0.15 |

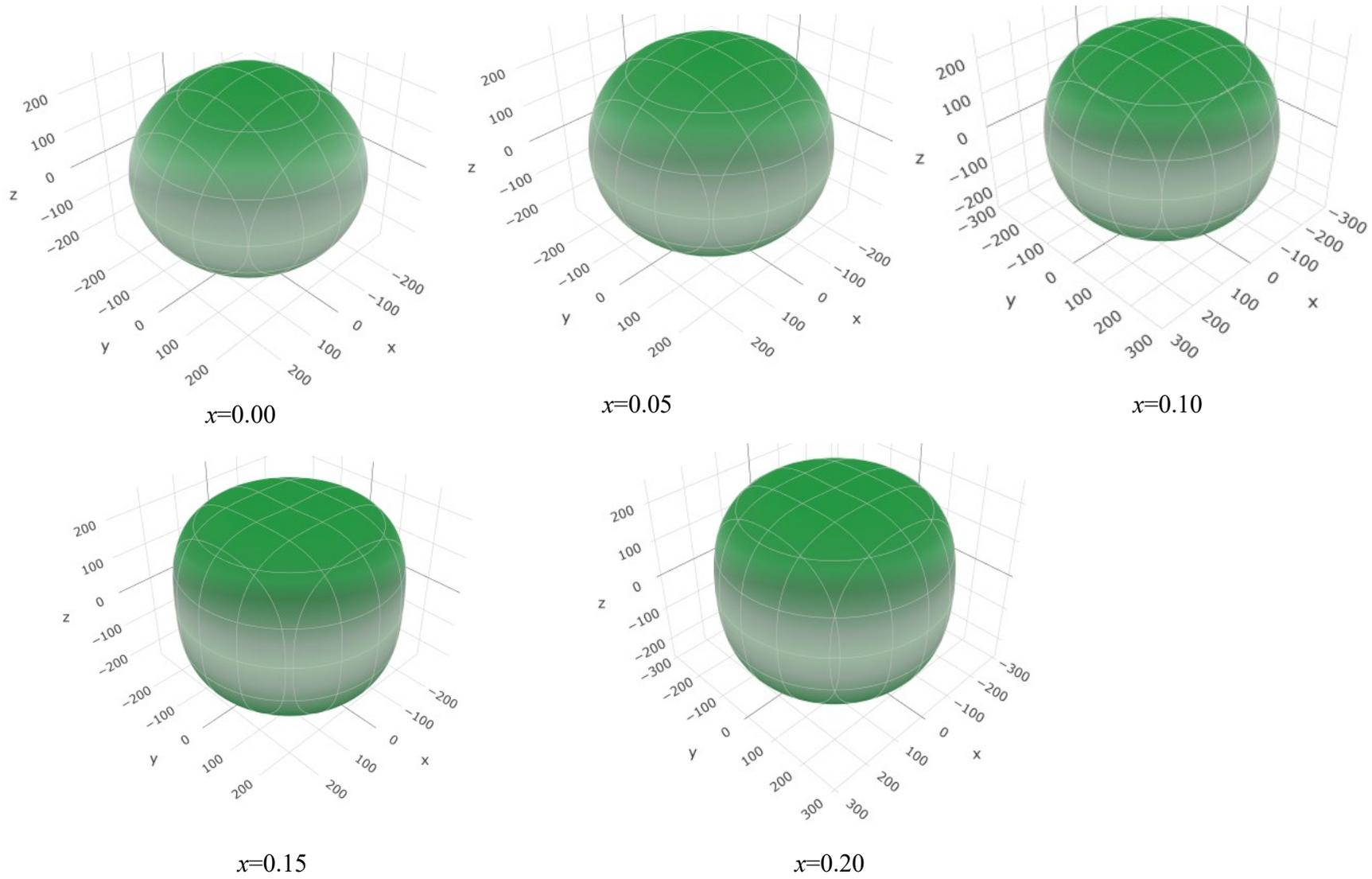

Fig. 5. The three-dimensional contour plots of the Young's modulus for $(Ti_{1-x}Mo_x)_2AlC$ solid solutions.

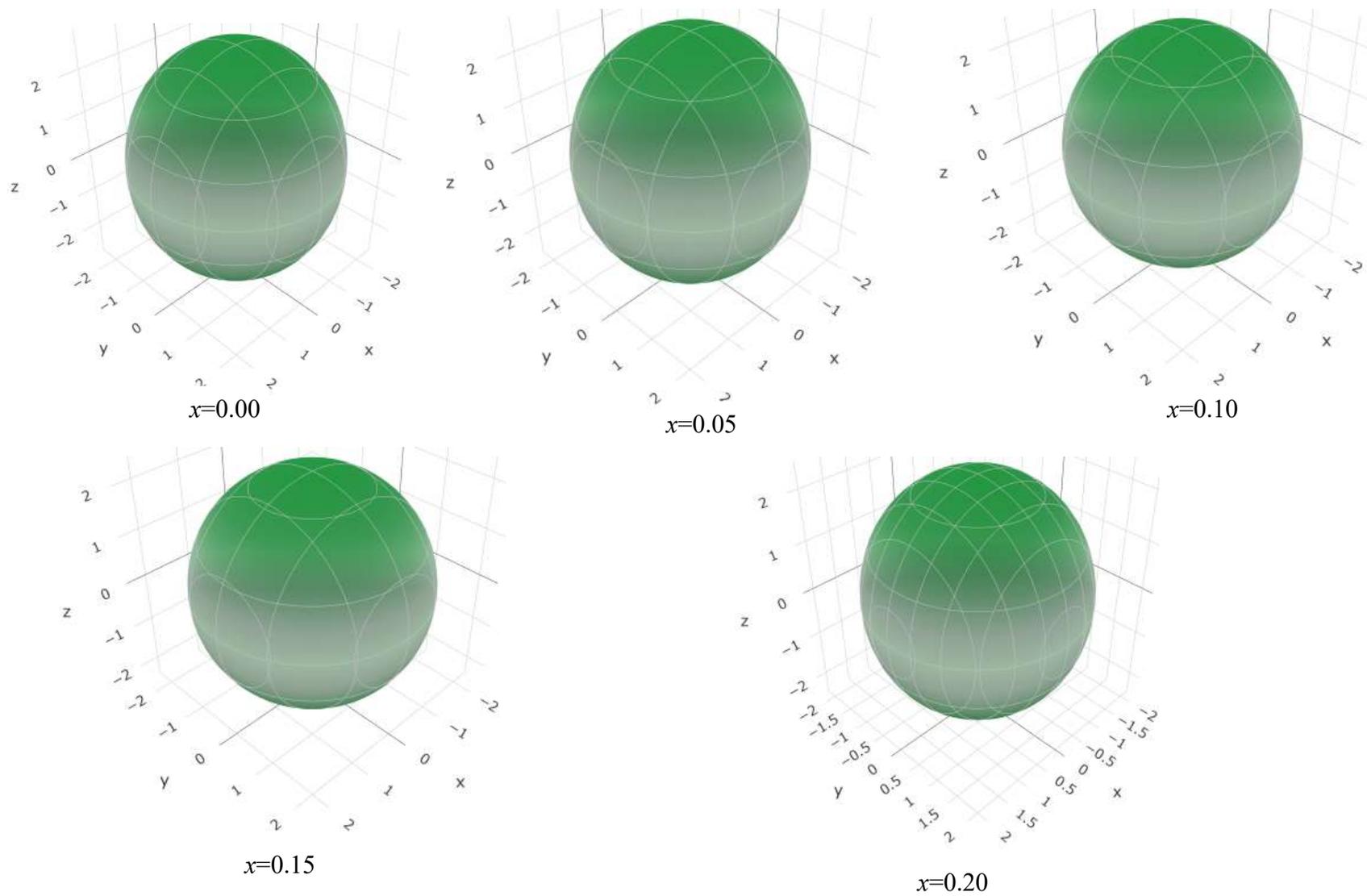

Fig. 6. The three-dimensional contour plots of the 1/bulk modulus for $(Ti_{1-x}Mo_x)_2AlC$ solid solutions.

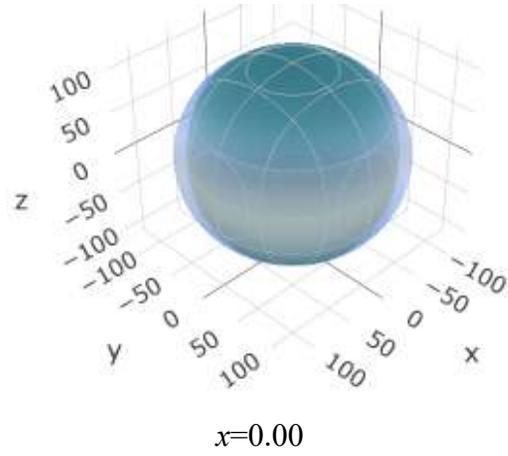

$x$=0.00

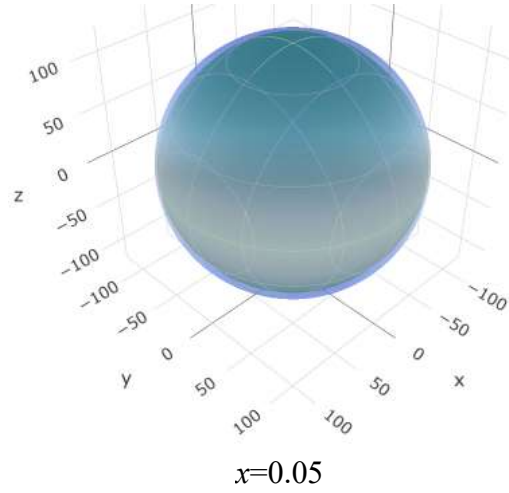

$x$=0.05

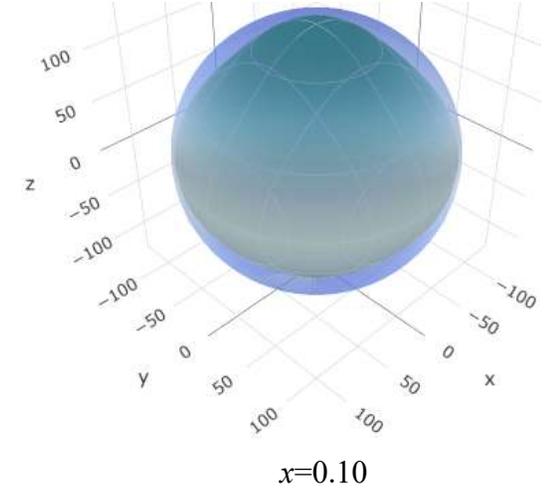

$x$=0.10

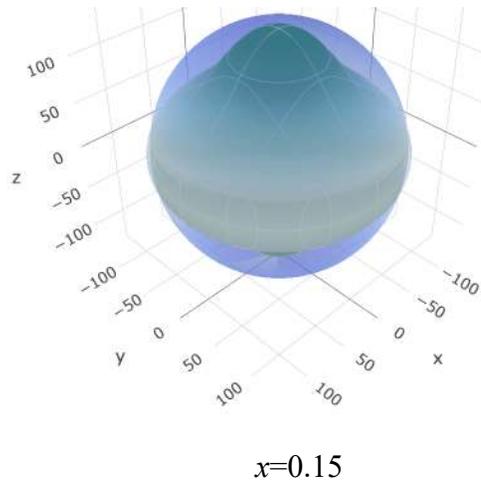

$x$=0.15

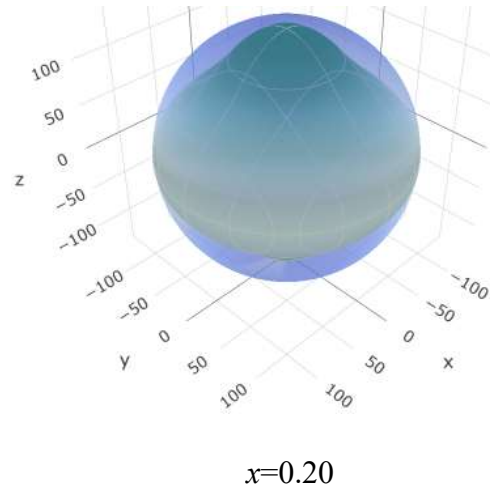

$x$=0.20

Fig. 7. The three-dimensional contour plots of the shear modulus for $(Ti_{1-x}Mo_x)_2AlC$ solid solutions.

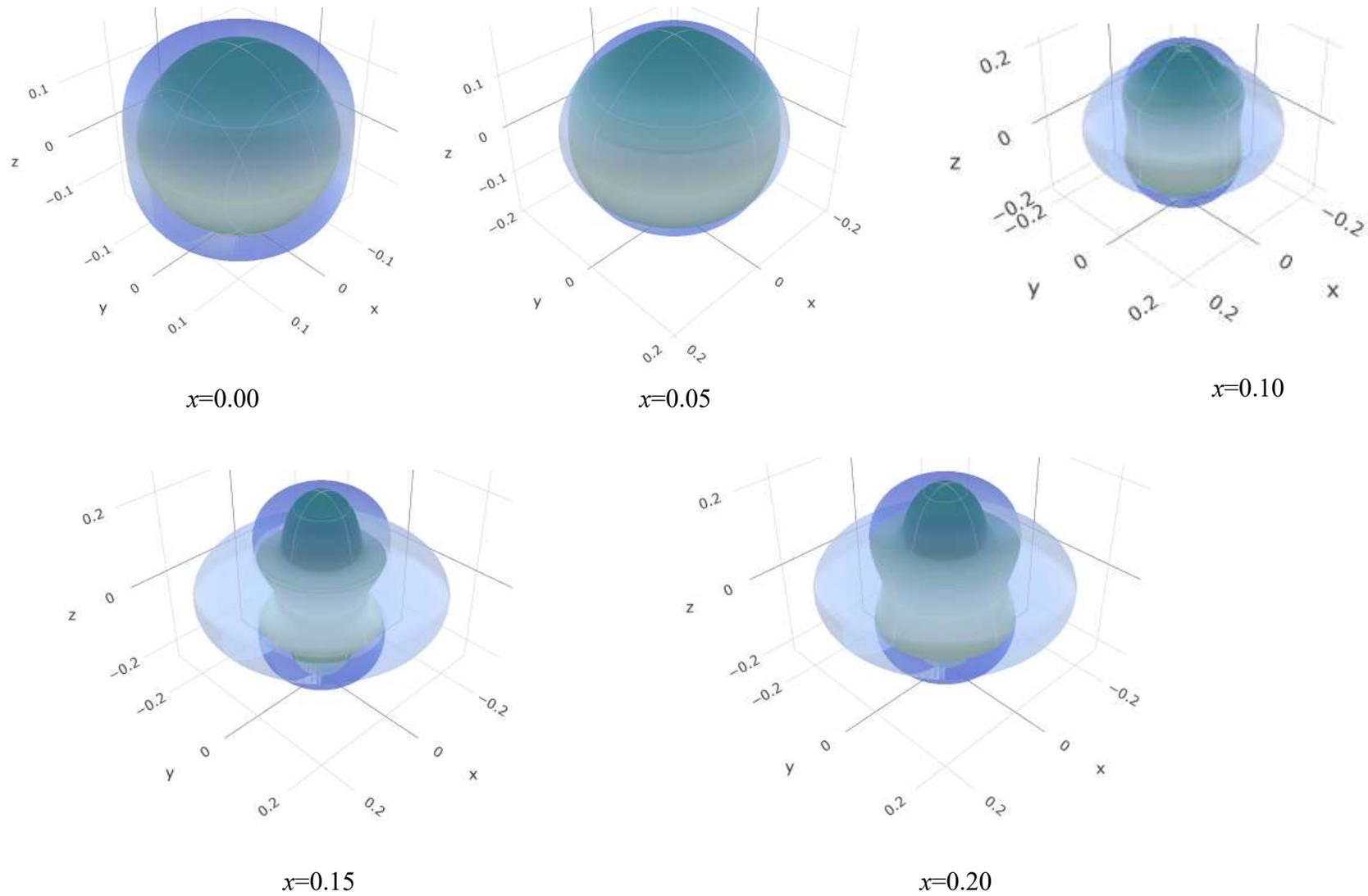

Fig. 8. The three-dimensional contour plots of the Poisson's ratio for (Ti$_{1-x}$Mo$_x$)$_2$AlC solid solutions.

### 3.4. *Thermodynamic properties*

The important thermodynamic properties such as Debye temperature, minimum thermal conductivity, Grüneisen parameter and melting temperature of the $(Ti_{1-x}Mo_x)_2AlC$ solid solutions are studied in this section.

One of the fundamental characteristics parameters of solids is the Debye temperature $(\Theta_D)$ which promotes understanding of a number of important physical parameters by correlating the mechanical properties with thermodynamic properties such as phonons, lattice vibrations, enthalpy, thermal conductivity, melting point, specific heat etc [94]. A simple but widely used method (details can be found elsewhere [95-97]) proposed by Anderson [98] has been used to calculate $\Theta_D$ using the average sound velocity in $(Ti_{1-x}Mo_x)_2AlC$ solid solutions. The sound velocities were determined from the elastic constants and crystal density of the compounds under consideration.

**Table 5**

Calculated density $(\rho)$, longitudinal, transverse and average sound velocities ($v_l$, $v_t$, and $v_m$, respectively), Debye temperature $(\Theta_D)$, minimum thermal conductivity $(K_{min})$, Grüneisen parameter $(\gamma)$ and melting temperature $(T_m)$ of $(Ti_{1-x}Mo_x)_2AlC$ solid solutions.

| $x$ | $\rho$ (g/cm³) | $v_l$ (m/s) | $v_t$ (m/s) | $v_m$ (m/s) | $\Theta_D$ (K) | $K_{min}$ (W/mK) | $\gamma$ | $T_m$ (K) |
|------|------|------|------|------|------|------|------|------|
| 0.00 | 3.997 | 8453 | 5316 | 5853 | 723 | 8.47 | 1.18 | 1663 |
|      | 4.013[a] | 8640[a] | 5350[a] | 5900[a] | 729[a] |  |  |  |
| 0.05 | 4.172 | 8446 | 5272 | 5808 | 719 | 8.45 | 1.21 | 1681 |
| 0.10 | 4.349 | 8356 | 5230 | 5761 | 715 | 8.43 | 1.20 | 1695 |
| 0.15 | 4.526 | 8248 | 5148 | 5672 | 706 | 8.34 | 1.21 | 1674 |
| 0.20 | 4.705 | 8238 | 5133 | 5656 | 705 | 8.35 | 1.22 | 1740 |

[a]Ref.-[72]

Table 5 displays the obtained values of sound velocities and $\Theta_D$ for different Mo content along with the previously reported values for $Ti_2AlC$ to assess the reliability of our calculations. The values of transverse sound velocity is much lower than the longitudinal sound velocity [99]; more energy is required to vibrate neighboring atoms during the propagation of transverse wave that results in the loss of wave energy and hence reduced the wave velocity. The calculated values of $\Theta_D$ presented in Table 5 are observed to vary in opposing trend of mechanical properties. As we know, in general, $\Theta_\text{D}$ are higher for harder solids [100, 101] and the hardness is assumed to be increased with increased elastic constants and moduli but the $\Theta_\text{D}$ is observed to be decreased with Mo contents. This apparent contradiction vanishes when we take into consideration of the much larger atomic weight of Mo (95.95) compared to Ti (46.86). In fact, within the harmonic approximation, the Debye frequency (equivalently the Debye temperature) varies inversely with the square-root of the atomic weight. Therefore, the slight reduction in the $\Theta_\text{D}$ with increasing Mo content does not imply that the lattice is becoming less stiff, it rather implies that the lattice is becoming relatively stiffer with Mo substitution when the huge difference in mass between Mo and Ti is properly taken into account. This follows from the harmonic approximation where the frequency ($\nu$) of the phonon mode can be expressed roughly as, $\nu \sim (k/M)^{1/2}$. Here, $k$ is the stiffness constant or force constant and $M$ is the mass of the atomic species.

The MAX phase materials are promising candidates for high temperatures applications. Therefore, study of high temperature physical properties is of scientific interest. Like, $\Theta_\text{D}$, another fundamental physical parameter is the lattice thermal conductivity ($K_\text{ph}$), resulting from role of phonon-phonon scattering on thermal transport. At high enough temperatures, the lattice thermal conductivity $K_\text{ph}$ assumes a minimum constant value ($K_\text{min}$). The phonons become completely uncoupled and the thermal energy is distributed evenly among the atoms within the solid [102]. We have calculated the $K_\text{min}$ using the following equation [103]: $K_{min} = k_B v_m \left( \frac{M}{n \rho N_A} \right)^{-2/3}$, where $k_\text{B}$, $v_\text{m}$, $N_\text{A}$ and $\rho$ are Boltzmann constant, average phonon velocity, Avogadro's number and density of crystal, respectively and presented those in Table 5. The values $K_\text{min}$ are observed to decrease very slowly with increasing Mo contents. This is largely due to the increase in the crystal density with increasing level of Mo substitution. The Grüneisen parameter ($\gamma$) provides the information regarding anharmonic effect in crystalline solids attributed from lattice dynamics. The Grüneisen parameter is calculated using Poisson's ratio: $\gamma = \frac{3}{2} \frac{(1+\nu)}{(2-3\nu)}$ [104]. The calculated values lie in the range of 0.85 to 3.53 for Poisson's ratio within the range 0.05–0.46 for diverse class of polycrystalline materials [105]. Relatively low and almost constant values of $\gamma$ of $(Ti_{1-x}Mo_x)_2AlC$ imply that anharmonic effects are low in all these alloys. The knowledge of melting temperature ($T_\text{m}$) of MAX phases is very important as they are potential candidates for high temperature applications. We have calculated the melting temperature of $(Ti_{1-x}Mo_x)_2AlC$ solid solutions via the empirical formula [106]: $T_m = 354 + \frac{4.5(2C_{11} + C_{33})}{3}$. The increased values of $T_m$ owing to Mo substitution indicates that compared to $Ti_2AlC$, Mo substituted alloys are more suitable for high temperature applications.

## 4. Conclusions

The Mo-substituted MAX phase solid solutions $(Ti_{1-x}Mo_x)_2AlC$ ($0 \leq x \leq 0.20$) have been investigated by density functional theory. From the detailed analysis of the results, the following conclusions can be drawn:

➤ Calculated lattice constants of $Ti_2AlC$, *a* and *c* are just 0.29% and 0.69%, respectively higher than the experimentally obtained values [71]. This shows very high level of agreement. The obtained lattice constants are also comparable with previously determined theoretical values of $Ti_2AlC$ [72-75].

➤ Nonexistence of energy band gap in the band structure and finite DOS (~ 3.0 states/eV) at Fermi level reveal the metallic nature of the studied solid solutions. Strongly anisotropic electrical conductivity is predicted in the $(Ti_{1-x}Mo_x)_2AlC$ solid solutions. The shifting of dispersion curves in energy in the band structure, shifting of peaks in the DOS towards lower energy and increased accumulation of charge carrier at Ti/Mo sites due to Mo substitution indicate that covalent bonds are strengthened between unlike atoms (metal-*d* and C-*p* states as well as metal-*d* and Al-*p* states) owing to Mo substitution should lead to enhanced mechanical properties.

➤ The Fermi level is located in the middle of bonding and antibonding peaks in the DOS curves (Fig. 2d) for $Ti_2AlC$ and $(Ti_{1-x}Mo_x)_2AlC$ alloys. This suggests that the solid solutions under investigation should possess high degree of electronic stability.

➤ The computed stiffness constants $(C_{ij})$ and elastic constants ($B$, $G$, $Y$, $G/B$ and $\upsilon$) are found in good accord with reported values (for $x = 0.0$) [56, 72, 74, 81-83]. The mechanical stability of the studied solid solutions is checked by Born criteria based on the $C_{ij}$. The single crystal elastic constants ($C_{11}$, $C_{33}$ and $C_{44}$) and other mechanical parameters of studied solid solutions are found to be increased to a considerable extent owing to Mo substitution for Ti and suggesting the enhanced mechanical hardness for the same. The values of $C_{11}$, $C_{33}$, $C_{44}$,

*B*, *G* and *Y* are increased by 6.5%, 3.3%, 22.4%, 14.0%, 9.7% and 10.5 %, respectively for $(Ti_{0.80}Mo_{0.20})_2AlC$.

➤ The Pugh ratio (greater than 0.571), Poisson's ratio (less than 0.26) and Cauchy pressure (-48 to -74) confirm the brittleness of studied compositions.

➤ The 3D contour plot confirmed the direction dependency of *B, G, Y* and υ. The shear anisotropic factors and universal anisotropy index also revealed the anisotropic nature of the considered solid solutions.

➤ The values of Debye temperature and minimum thermal conductivity are obsreved to decrease ($\Theta_D$: from 723 K for *x* = 0.00 to 705 K for *x* = 0.20) while the Grüneisen parameter and melting temperature (*γ:* from 1.18 to 1.22 and $T_m$: from 1663 K to 1740 K for *x* = 0.00 and *x* = 0.20, respectively) are noted to be increased due to Mo substitution. The Debye temperature for $(Ti_{1-x}Mo_x)_2AlC$ alloys are quite high compared to many other MAX phase nanolaminates [107-110]

To summarize, this study has found that Mo substitution can lead to significant improvement of the mechanical and thermal properties of $(Ti_{1-x}Mo_x)_2AlC$ solid solutions. We hope that the scientific community will be inspired to synthesize and study the Mo substituted solids solutions of MAX phases belonging to different sub-classes in future.

## Data availability

The data sets generated and/or analyzed in this study are available from the corresponding authors on reasonable request.

**Author Contributions**

M. A. A. designed the project, performed the analysis and wrote the draft manuscript. S. H. N. finalized the draft paper and supervised the project.

**Additional Information**

**Competing Interests**

The authors declare no competing interests.